\documentclass{pnastwo}
\usepackage{graphicx}
\usepackage{amssymb,amsfonts,amsmath}
\usepackage{cite}
\usepackage{color}

\begin{document}

\title{The Role of Projection in the Control of Bird Flocks}

\author{Daniel J.G. Pearce\affil{1}{Department of Physics}\affil{2}{MOAC Doctoral Training Centre},
A. M. Miller\affil{1}{}\affil{3}{Centre for Complexity Science, University of Warwick, Coventry CV4 7AL, United Kingdom},
George Rowlands\affil{1}
\and
Matthew S. Turner\affil{1}{}\affil{3}{Centre for Complexity Science, University of Warwick, Coventry CV4 7AL, United Kingdom}\affil{4}{Laboratoire Gulliver (CNRS UMR 7083), ESPCI, 10 rue Vauquelin, 75231 Paris Cedex 05, France}
}

\maketitle

\begin{article}
\begin{abstract}
Swarming is a conspicuous behavioural trait observed in bird flocks, fish shoals, insect swarms and mammal herds. It is thought to improve collective awareness and offer protection from predators. Many current models involve the hypothesis that information coordinating motion is exchanged between neighbors. We argue that such local interactions alone are insufficient to explain the organization of large flocks of birds and that the mechanism for the exchange of long-ranged information necessary to control their density remains unknown. We show that large flocks self-organize to the maximum density at which a typical individual is still just able to see out of the flock in many directions.  Such flocks are marginally opaque - an external observer can also just still see a substantial fraction of sky through the flock. Although seemingly intuitive we show that this need not be the case; flocks could easily be highly diffuse or entirely opaque. The emergence of marginal opacity strongly constrains how individuals interact with each other within large swarms. It also provides a mechanism for global interactions: An individual can respond to the projection of the flock that it sees. This provides for faster information transfer and hence rapid flock dynamics, another advantage over local models. From a behavioural perspective it optimizes the information available to each bird while maintaining the protection of a dense, coherent flock.
\end{abstract}


\dropcap{S}tarling murmurations represent one of the most impressive examples of organization in the natural world, with flocks of up to 300,000 individuals or more able to coordinate themselves into a cohesive and highly coherent group \cite{Ballerinistudy, Murm,pigeoncognition,birdbrainbook, ScaleFree}.

While the primary source of sensory information to a bird is visual, it would be unrealistic to expect that individual to recognize and track the position and orientation of a significant proportion of the other members of a flock \cite{birdbrainbook,pigeoncognition}. Indeed, observations on real starling flocks show that a bird only responds to this type of information from its 7 nearest neighbors and that these interactions are scale free \cite{Ballerinistudy,ScaleFree,ballerinitopol}. Local interactions like this are enough to create order within a flock \cite{ballerinitopol,ScaleFree,vicsek,chatevoronoi,visualnetw,bialekStatmec}, but do not give any information on the state of the flock as a whole nor do they explain how density might be regulated. Most models employ attraction and repulsion interactions, a fictitous potential field or simply fix the available volume in order to control the density\cite{chateAR,swarmmodel,OrsognaAR,couzinAR,giardinarev,ARmodelreview,HHmodel,HHmodel2012,vicsek,chatevoronoi,ballerinitopol,vicsekCOM,gregprl}. 

In order to make progress we first ask a simple question,``What does a bird {actually} see when it is part of a large flock?'' Its view out from within a large flock would likely present the vast majority of individuals as merely silhouettes, moving too fast and at too great a distance to be easily tracked or even discriminated from each other. Here the basic visual input to each individual is assumed to be simply based on visual contrast: a dynamic pattern of dark (bird) and light (sky) across the field of vision (although it might be possible to extend this to other swarming species and environmental backgrounds, respectively). This has the appealing feature that it is also the projection that appears on the retina of the bird, which we assume to be its primary sensory input. A typical individual within a very dense flock would see other, overlapping individuals (dark) almost everywhere it looked. Conversely, an isolated individual, detached from the flock, would see only sky (light). The projected view gives direct information on the {\it global} state of the flock. It is a lower dimensional projection of the full 6$N$ degrees of freedom of the flock and is therefore more computationally manageable, both for the birds themselves and for the construction of simple mathematical models of swarm behaviour. 

The information required to mathematically specify the projection is linear in the number of boundaries. Our simplifying assumption is that the individual only registers such a black-and-white projection (in addition to nearest neighbour orientation). This is then all the information that would be available to an agent, regardless of the behavioural model that might be chosen. Individuals in a flock that is sufficiently sparse for them to typically see a complex projected pattern of dark and light have more information about the global state of the flock. Such sparse flocks also allow an individual to see out in a significant fraction of all directions, which would allow the approach of a predator, or at least the response of distant individuals to the approach of predator, to be registered. Conversely a dense, completely opaque flock would offer little information, either about the global state of the flock or the approach of predators.

We define the opacity, $\Theta'$, of a flock to be the fraction of sky obscured by individuals \textit{from the viewpoint of a distant external observer}. A closely related quantity is the average opacity seen by a typical individual \textit{located within the flock}, written $\Theta$. Crucially, the opacity and density are quite different quantities: Flocks containing large 
numbers of individuals could be nearly opaque ($\Theta \approx 1$) even for very small densities, corresponding to well separated birds. Below we present evidence that large bird flocks are {\it marginally opaque}, with opacities that are intermediate, neither very close to 0 nor 1 ($0.25\lesssim\Theta'\lesssim 0.6$ in our data). Such a state corresponds to a complex projected pattern rich in information.

In the remainder of this article we will focus on proposing a model for {\it how} bird flocks organise and specifically on how the global density is regulated, which remains an open question \cite{Ballerinistudy}. We will develop what we believe to be the simplest possible model that takes the projected view described above as sensory input, while retaining co-alignment with (visible) nearest neighbours and allowing for some noise. We then compare the swarms generated by this model with data.

\subsection{Hybrid Projection Model}   

\begin{figure*}[ht]
\centerline{\includegraphics[width=0.4\textwidth]{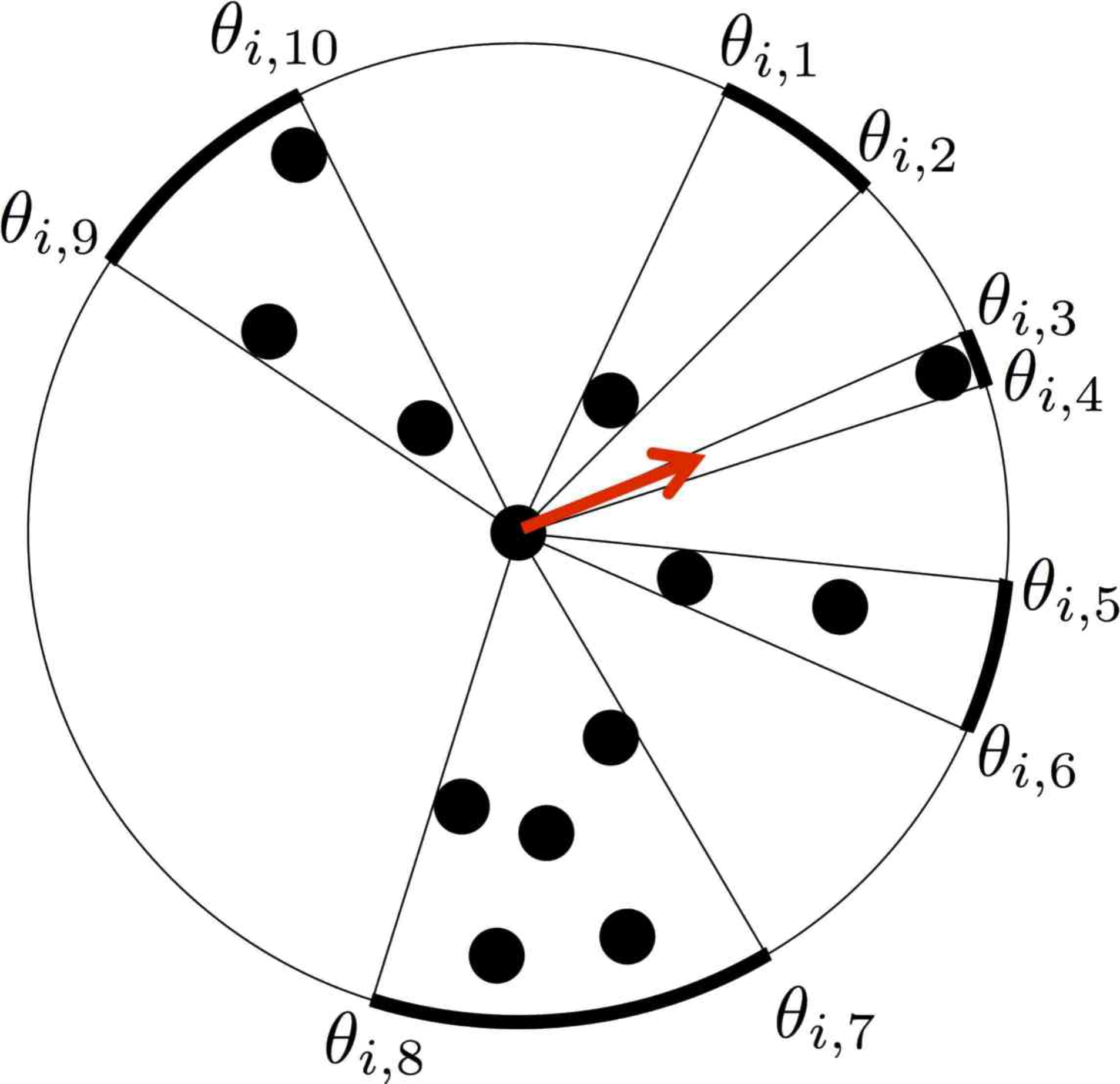}}
\caption{Sketch showing the construction of the projection through a 2D swarm seen by the $i^{\rm th}$ individual, which here happens to be one near the centre of the swarm. The thick, dark arcs around the exterior circle (shown for clarity - there is no such boundary around the swarm) correspond to those angular regions where one or more others block the line of sight of the $i^{\rm th}$ individual to infinity. The sum of unit vectors pointing to each of these domain boundaries, at the angles shown, gives the resolved vector $\underline{\delta}_{i}$, shown in red, that enters our equation of motion. {See SI for the extension to 3D}.}
\label{fig:deltafig}
\end{figure*}

We propose a \textit{hybrid projection} model in which each individual responds to the projection through the swarm that it observes.  We first identify those (dark) angular regions where a line-of-sight traced from an individual to infinity intersects one {\it or more} other members of the swarm. These are separated by (light) domains, see Fig.~\ref{fig:deltafig}. 

Each individual is assumed to be isotropic and has a size $b=1$, that then defines our units of length. Anisotropic bodies give rise to a projected size that depends on orientation and are explored further in the SI. In two dimensions the domain boundaries seen by the $i^{\rm th}$ individual define a set of angles $\theta_{ij}$, measured from an arbitrary reference ($x$) axis, where the index $j$ runs over all the $\mathcal{N}_i$ light-dark (or dark-light) domain boundaries seen by the $i^{\rm th}$ individual, equal to 10 for the central individual shown in Fig.~\ref{fig:deltafig}. These $\theta_{ij}$ are a reasonable choice for input to a behavioural model: edge detection like this is known to be performed in the neural hardware of the visual cortex in higher animals\cite{geisler1997visual}. In particular behavioural models based on motional bias towards either the most dark or light regions tend to be unstable with respect to  collapse or expansion respectively. The most simple candidate model that might support physically reasonable solutions is therefore one that responds to the domain boundaries. We seek a model that takes as input the angles specifying the domain boundaries and produces a characteristic direction for the birds, acknowledging that their actual motion should also include their known tendency to co-align with neighbours and also the effect of some noise. A natural choice for this characteristic direction is simply the average direction to all boundaries $\underline{\delta}_{i}$
\begin{equation}
\underline{\delta}_{i}=\frac1{\mathcal{N}_{i}}\sum_{j=1}^{\mathcal{N}_i}\left({\cos \theta_{ij}}\atop{\sin \theta_{ij}}\right)
\label{delta}
\end{equation}

\noindent {This can easily be extended to three dimensional flocks, where the light dark boundaries can now be represented as curves on the surface of a sphere and $\delta$ becomes the normalised integral of radial unit vectors traced along these curves, see SI for details.}

Our model will involve $\underline{\delta}_{i}$ in such a way as to correspond to birds being equally attracted to all the light-dark domain boundaries. In addition they co-align with {\it visible} local neighbors, assigned in a topological fashion \cite{ballerinitopol,visualnetw}. We define visible neighbours to be those for which there is an unbroken line of sight between the two individuals (see SI for details). We incorporate these two preferred directions, arising from the projection and the motion of neighbours, into an otherwise standard agent-based model for a swarm of $N$ particles moving off-lattice with constant speed $v_{0}$ ($v_0=1$ in all our simulations). For simplicity we treat the individuals as ``phantoms", having no direct steric interactions (the effect of introducing steric interactions is explored further in the SI). The equation of motion for the position $\underline{r}_{i}^{t}$ of the $i^{\rm th}$ individual at discrete time $t$ is

\begin{equation}
\underline{r}_{i}^{t+1} = \underline{r}_{i}^{t} + v_{0} \hat{\underline{v}}_{i}^{t}
\label{rupdate}
\end{equation}
with a velocity parallel to
\begin{equation}
\underline{v}_{i}^{t+1}=\phi_{p}\underline{\delta}_{i}^t +\phi_{a}\widehat{ \langle\underline{v}_{k}^{t}\rangle}_{n.n.} + \phi_{n}\underline{\eta}_i^t
\label{vupdate}
\end{equation}
where $\langle\dots\rangle_{n.n.}$ is an average over the $k\in [1,\sigma]$ nearest neighbours to the $i^{\rm th}$ individual ($\sigma=4$ in all simulations), a hat, $\hat{}$, denotes a normalised vector and $\underline{\eta}_i^t$ is a noise term of unit magnitude having a different (uncorrelated) random orientation for each individual at each timestep. This equation involves only three primary control parameters $\phi_{p}$, $\phi_a$ and $\phi_n$, the  weights of the projection alignment and noise terms, respectively. We further simplify by considering only the relative magnitudes (ratios) of these control parameters which are then taken to obey
\begin{equation}
\phi_{p}+\phi_{a}+\phi_{n}=1 
\label{conservation}
\end{equation}
We now analyse the results of computer simulation of the swarms arising from these equations of motion for given combinations of $\{\phi_p,\phi_a\}$ alone, with $\phi_n$ given by construction through Eq~(\ref{conservation}).  Several distinct behavioural phenotypes reminiscent of birds, fish and insects are observed (see SI movies 5, 6 \& 7 respectively). Further generalisations of the model are also explored in the SI, including the effect of steric/repulsive interactions and incomplete angular vision corresponding to ``blind angles'' behind each bird.

\subsection{The Hybrid Projection Model reproduces key features of a flock of birds}
\begin{figure*}[ht]
\centerline{\includegraphics[width=0.7\textwidth]{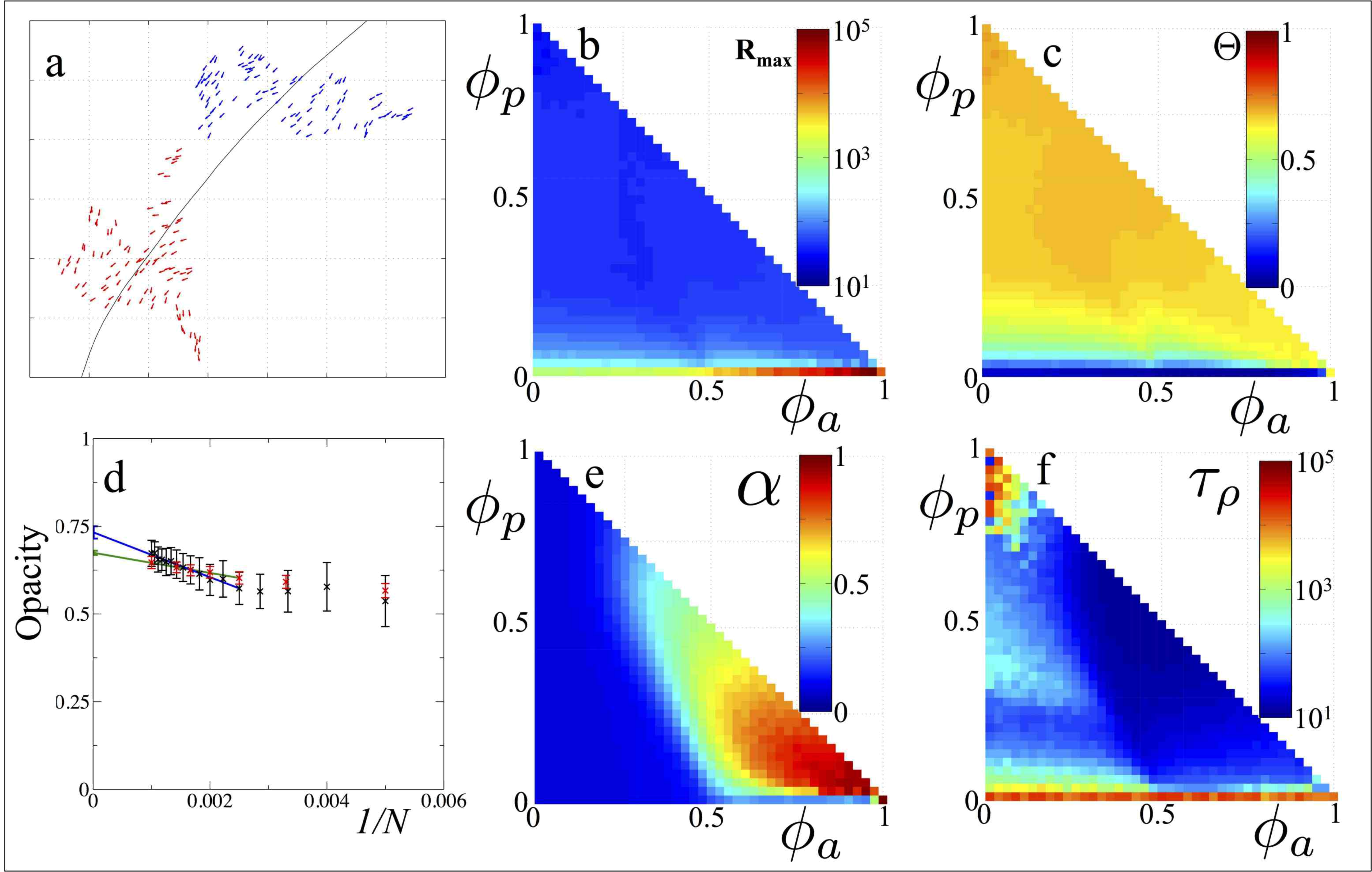}}
\caption{Results from repeated computer simulation of a simple hybrid projection model, parameterized by the strength of the response of each individual to the projection through the swarm that they see ($\phi_p$) and the strength of the alignment with their 4 nearest neighbors ($\phi_a$). In (b), (c), (e) \& (f) (2D) each small colored square (point), corresponding to a pair of parameter values $\{ \phi_a,\phi_p \}$, is an average value over 400,000 timesteps for $N=100$ individuals. (a) A snapshot of a swarm in 2D with $\phi_a=0.75$ and $\phi_p=0.1$ at two different times (blue then red). Its centre of mass is moving along the solid line. (b) The distance between the two furthest individuals in the swarm, $R_{max}$, in units of particle diameter - the swarm does not fragment unless $\phi_p=0$. (c) The average opacity $\Theta$. (d) The average opacity of swarms containing different numbers of individuals $N$ (the axis shows $1/N$), as seen by internal observers for 2D (black) and 3D (red) swarms, with $\phi_p=0.03$ and $\phi_a=0.8$ averaged over at least 50,000 timesteps. {The linear fit with an $R^{2}$ value of 0.97 for 2D (blue) and 0.99 for 3D (green), is to all data points $N\ge 400$.} (e) The average speed, $\alpha$, of the centre of mass of the swarm, normalized by the individualÕs speed. This is sometimes refereed to as the order parameter. (f) The swarm density autocorrelation time $\tau_\rho$ in simulation time-steps. {The upper left corner of this panel represents dynamically ``jammed" states that we believe to be unphysical (see SI for details).}}\label{Simdata}
\end{figure*}

In particular it naturally leads to robustly cohesive swarms, see Fig.~\ref{Simdata} a \& b (and the SI) as well as the emergence of \textit{marginal opacity} in large flocks of birds where  both $\Theta$ and $\Theta'$ are neither very close to 0 nor 1, see Fig.~\ref{Simdata} c \& d. 

The emergence of marginal opacity is a new feature and it is worth emphasizing that the model was not constructed so as to target any particular ``preferred'' opacity value, rather marginal opacity emerges naturally. {Importantly, it arises for swarms of varying size $N$ that are realised with exactly the same control parameters $\phi_p$ and $\phi_a$. This means that marginal opacity can be maintained without a bird changing its behaviour with, or even being aware of, the size of the flock.} Other models, which control the density in a metric fashion \cite{chateAR,couzinAR}, give rise to values for $\Theta$ that approach 1  {as the number of individuals in a swarm increases,} i.e. they always become fully opaque (see SI for details).  {In such metric-based models the density of the swarm is fixed by the control parameters. Thus, for any combination of these parameters there will always be a critical size at which the swarm becomes opaque. For the typical values analysed in the literature rather small flocks with $N<100$ are already fully opaque (see SI for details). The only possible approach to preventing swarms from becoming opaque with such models would be to continuously modify their control parameters as a function of the swarm size. This would represent a significant proliferation in control parameters from a baseline level that is already typically far higher than in the present work. This is the signature of a class of models that are structurally inadequate to explain marginal opacity. }

\begin{figure*}[ht]
\begin{center}
\centerline{\includegraphics[width=.7\textwidth]{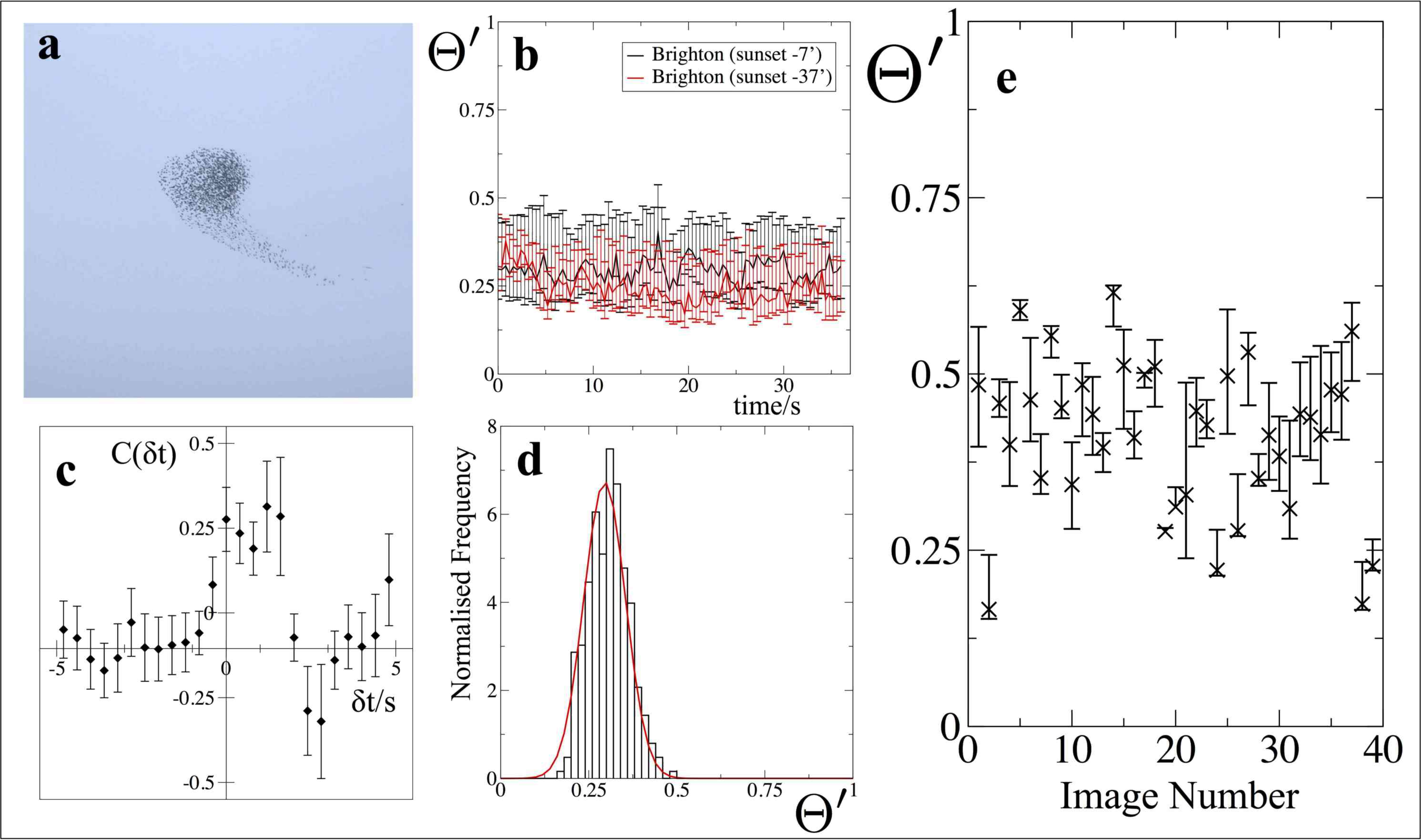}}
\caption{(a) A snapshot of a flock of Starlings (image contributes to the data presented in panels b-d). (b) Typical time variation of the opacity $\Theta'$ of starling flocks observed in dim light (black) and under brighter conditions (red). (c) Cross-correlation function of the horizontal acceleration $a$ of the centre of mass of a flock and its opacity $C(\delta t)$ as a function of the delay $\delta t$. (d) Histogram of the opacity $\Theta'$ of different Starling flocks from across the UK corresponding to $n=118$ uncorrelated measurements. The red line displays a gaussian distribution fitted to this data with $\mu = 0.30, \sigma^2 = 0.059$. (e) The opacity $\Theta'$ of images of starling flocks in the public domain ($\mu = 0.41, \sigma^2 = 0.012$). {In both (d) and (e) the null hypothesis that the opacities are drawn from a uniform distribution on [X,1] can be rejected at the 99.99\% confidence level for all values of X. These flocks are all {\it marginally opaque}.} See SI for details throughout.}\label{Birddata}
\end{center}
\end{figure*}

{In Fig.~\ref{Simdata} b,c,e,f we see that individuals don't respond to the projection at all in the narrow strip where $\phi_p=0$}. Here the swarm fragments/disperses. Provided there is even a very weak coupling to the projection, i.e. $\phi_p>0$, the swarm no longer dissipates (see also SI). In Fig.~\ref{Simdata}f the narrow red strip near $\phi_p=0$ shows that the response of the swarm is slow in the absence of the projection term. Here, even when the swarm does not fragment, the dynamics depend on the exchange of information between nearest neighbors. The correlation time decreases as the strength of response to the projection is turned on. This is because the projection provides a global interaction and can therefore lead to rapid dynamic response, consistent with the fast transients observed in real flocks. The nature of this model also makes it robust in response to shocks, suck as those caused by predation in real animal systems (see SI Movie 3). We now compare our model with data on flocks of starlings, see Fig.~\ref{Birddata}. 

Datasets for the 3D positions of birds in a flock, such as reported in \cite{ScaleFree,Ballerinistudy}, have given us many new insights but there are well known issues associated with particle tracking techniques in high density flocks. This makes using these techniques to obtain unbiased measurements of opacity itself problematic. Instead we chose to study data for 2D projections, this being best suited to test our prediction of projected opacity. Fig.~\ref{Birddata}b shows that the opacity remains roughly constant over a period in which the flock reversed direction several times. Fig.~\ref{Birddata}c shows that opacity changes significantly within a few seconds of rapid acceleration and could therefore be implicated in long-ranged information exchange across the flock. The crucial feature in both Fig.~\ref{Birddata}d (our data) and Fig.~\ref{Birddata}e (public domain images) is that the opacity is intermediate, i.e. neither very close to zero nor unity, in spite of the fact that the flocks had very different sizes and were observed under different conditions (the flocks we analyzed in (Fig.~\ref{Birddata}b-d) are generally smaller than in (Fig.~\ref{Birddata}e)). This is a feature that, to our knowledge, is not found in any existing models but emerges naturally from our hybrid projection model. 

It is insightful to consider the following simple mean field argument for the consequences of marginal opacity: Consider a randomly chosen line of sight through, or out from a typical location near the centre of, an idealized homogenous, isotropic flock. Because the probability that any small region is occupied is proportional to its volume multiplied by the density of individuals, the probability that a line of sight reveals ``sky'' is Poisson distributed according to $P_{sky} \approx e^{-\rho b^{d-1} R}$ with $\rho = N/R^{d}$ a d-dimensional density, $b$ the effective linear size of an individual and R the linear size of the flock. Our hypothesis of marginal opacity corresponds to $P_{sky}$ being of order unity (a half, say) leading to $\rho \sim N^{-1/(d-1)}$, i.e. $\rho \sim N^{-1}$ in 2D and $\rho \sim N^{-1/2}$ in 3D. Marginal opacity therefore either requires the density to be a decreasing function of $N$ or for the flock morphology to change (or both). There are hints of both of these qualities in some published data \cite{ScaleFree,Ballerinistudy} not inconsistent with the predictions of our model. 

Our mean field analysis can also be used to understand why the emergence of marginal opacity is quite such a surprising result. It follows that most spatial arrangements of $N$ finite sized particles are either opaque ($\Theta \approx 1$) or predominantly transparent ($\Theta\ll1$). The latter obviously occurs whenever the density is very low (and in an essentially infinite space there is plenty of room to achieve this) while the former arises even for a relatively small reduction in the separation between individuals from that found in the marginal state. This is due to the extremely strong dependence of $P_{sky}$ on the flock size $R$ (in 3D it varies {\em exponentially} with the {\em square} of $R$).
To illustrate this we consider the effect of a reduction by half of the spacing between individuals, and hence also $R$. Using  $P_{sky} \approx e^{-N(b/R)^{2}}$ in 3D we find that this leads to a change in opacity from (say) 50\% before to 94\% afterwards. Thus, the flock becomes almost completely opaque as a consequence of only a halving of the inter-bird spacing. 
{Similar arguments apply if $N$ increases at constant $R$ and such variations in both density and size are reported in the literature (e.g. \cite{Ballerinistudy}, table 1), supporting the claim that the marginal opacity apparent in Fig.~\ref{Birddata}b,d \& e is a robust emergent feature}.

We believe that opacity may be related to evolutionary fitness in flocking animals. Dense swarms are thought to give an advantage against predation due to {\it target degeneracy} in which the predator has difficulty distinguishing individual targets \cite{miller1922}. Balancing this is the need for the individuals to be aware of the predator, so as to execute evasion. In flocks with very high opacity only a very small fraction of all individuals would be able to see out of the flock and monitor either the first or subsequent approaches of the predator. Individuals in the interior of such a flock would neither be able to see the predator directly nor respond to the behaviour of individuals near the edge that were able to see it. Information about the approaches of a predator would instead have to propagate inwards, being passed from (the behaviour of) neighbour to neighbour, i.e. very much slower than the speed of light, which would instead operate on a clear line of sight. The state of marginal opacity would therefore seem to balance the benefit of compactness (target degeneracy) with information (``many eyes"\cite{pulliam1973}). In particular, there would be very little gain in infomation from decreasing the opacity beyond a marginal state. Thus, projection-based models that give rise to marginally opaque states would seem to be both cognitively plausible and evolutionarily fit.

Modern humans also need to rapidly extract useful information from high dimensional datasets. A generic approach to this is to present information through lower dimensional projections. This is reminiscent of the approach that we are proposing has been adopted by flocking animals. Here a 2$dN$-dimensional phase space, consisting of the spatial coordinates and velocities of all $N$ members of the flock, is projected onto a simple pattern on a line (2D) or surface (3D). Perhaps the use of such simplifying projections is more widespread in nature than previously suspected?

\begin{acknowledgments}
This work was partially supported by the UK Engineering and Physical Sciences Research Council through the MOAC and Complexity Doctoral Training Centres (to DJGP and AMM, respectively) and grant EP/E501311/1 (a Leadership Fellowship to MST). MST is also grateful for a Joliot-Curie visiting professorship at ESPCI Paris and the generous hospitality of the Physico-Chimie Th\'eorique group.

\end{acknowledgments}

\end{article}

\end{document}